\documentclass[%
 aip,
 jap,%
 amsmath,amssymb,
reprint,%
]{revtex4-1}
\usepackage{graphicx}
\usepackage{dcolumn}
\usepackage{bm}
\usepackage{color} 
\usepackage{multirow}
\usepackage{balance}
\usepackage[update,prepend]{epstopdf} 
\makeatletter

\newcommand{\Rmnum}[1]{\expandafter\@slowromancap\romannumeral #1@}
\makeatother
\begin{document}
\preprint{AIP/123-QED}
\title{Antisite Disorder-induced Exchange Bias Effect in Multiferroic Y$_2$CoMnO$_6$}
\author{Harikrishnan S. Nair}
\affiliation{Highly Correlated Matter Research Group, Physics Department, University of Johannesburg, P.O. Box 524, Auckland Park 2006, South Africa}
\affiliation{J\"{u}lich Center for Neutron Sciences and Peter Gr\"{u}nberg Institute, JARA-FIT, Forschungszentrum J\"{u}lich GmbH, 52425 J\"{u}lich, Germany}
\author{Tapan Chatterji}
\affiliation{Institut Laue-Langevin, BP 156, 38042 Grenoble Cedex 9, France}
\author{Andr\'{e} M. Strydom}
\affiliation{Highly Correlated Matter Research Group, Physics Department, University of Johannesburg, P.O. Box 524, Auckland Park 2006, South Africa}
\affiliation{Max Planck Institute for Chemical Physics of Solids (MPICPfS), N\"{o}thnitzerstra{\ss}e 40, 01187 Dresden, Germany}
\date{\today} 
%
\begin{abstract}
Exchange bias effect in the ferromagnetic double perovskite compound Y$_2$CoMnO$_6$, which is also a multiferroic, is reported. The exchange bias, observed below 8~K, is explained as arising due to the interface effect between the ferromagnetic and antiferromagnetic clusters created by {\it antisite} disorder in this material. Below 8~K, prominent ferromagnetic hysteresis with metamagnetic "steps'' and significant coercive field, $H_c \approx$ 10~kOe are observed in this compound which has a $T_c \approx$ 75~K.  A model based on growth of ferromagnetic domains overcoming the elastic energy of structurally pinned magnetic interfaces, which closely resembles martensitic-like transitions, is adapted to explain the observed effects.  The role of {\it antisite} disorder in creating the domain structure leading to exchange bias effect is highlighted in the present work.
\end{abstract}
\pacs{75.50.Lk, 75.47.Lx, 75.50.-y}
\maketitle
Exchange bias is a technologically important effect which is extremely useful in read-heads in magnetic recording, \cite{zhang_ieee_38_1861_2002magnetic} giant 
magnetoresistive random access memory devices, \cite{tehrani_ieee_91_703_2003magnetoresistive} in stabilizing nanoparticles \cite{nogues_prl_97_157203_2006shell} and in permanent magnets.\cite{sort_apl_79_1142_2001coercivity} It was first reported for fine particles of Co surrounded by CoO layers.\cite{meiklejohn_1956} Since its discovery, the effect has been observed in many materials and the importance of this area of research is documented in noted reviews. \cite{nogues_1999,giri_jpcm_23_073201_2011exchange} Exchange bias (EB) is interpreted as a phenomena occurring at the interface between magnetically ordered microscopic regions in a sample. These interfaces can be ferromagnetic/antiferromagnetic or ferromagnetic/antiferromagnetic/spin-glass and so on. Normally observed in tiny coated particles, nano-materials, thin films and inhomogeneous materials, reports on EB in bulk samples of correlated oxide systems are limited in number \cite{giri_jpcm_23_073201_2011exchange,pradheesh_apl_101_142401_2012exchange} though exchange bias effects observed in bulk materials like Heusler alloys have been of interest.\cite{khan_apl_91_072510_2007exchange,li_apl_91_112505_2007observation,khan_jap_102_113914_2007exchange} In this Letter we report on the observation of exchange bias in the ferromagnetic double perovskite Y$_2$CoMnO$_6$. This double perovskite is truly multifunctional in properties since, multiferroicity\cite{sharma_apl_103_012903_2013}, metamagnetic-like magnetization steps\cite{nair_jap_2014,murthy_arxiv2014} and multicaloric effects\cite{murthy2014multicaloric} have been reported on it so far. Our observation of exchange bias adds one more important functionality to this material. In the present work, results from field cooled (FC) and zero-field cooled (ZFC) magnetic hysteresis, training effect and ac susceptibility measurements are combined to demonstrate exchange bias effect  in this material. It is suggested that the observed exchange bias originates from {\it antisite} disorder commonly observed in double perovskites.
\\
\begin{figure}[!t]
\centering{}
\includegraphics[scale=0.22]{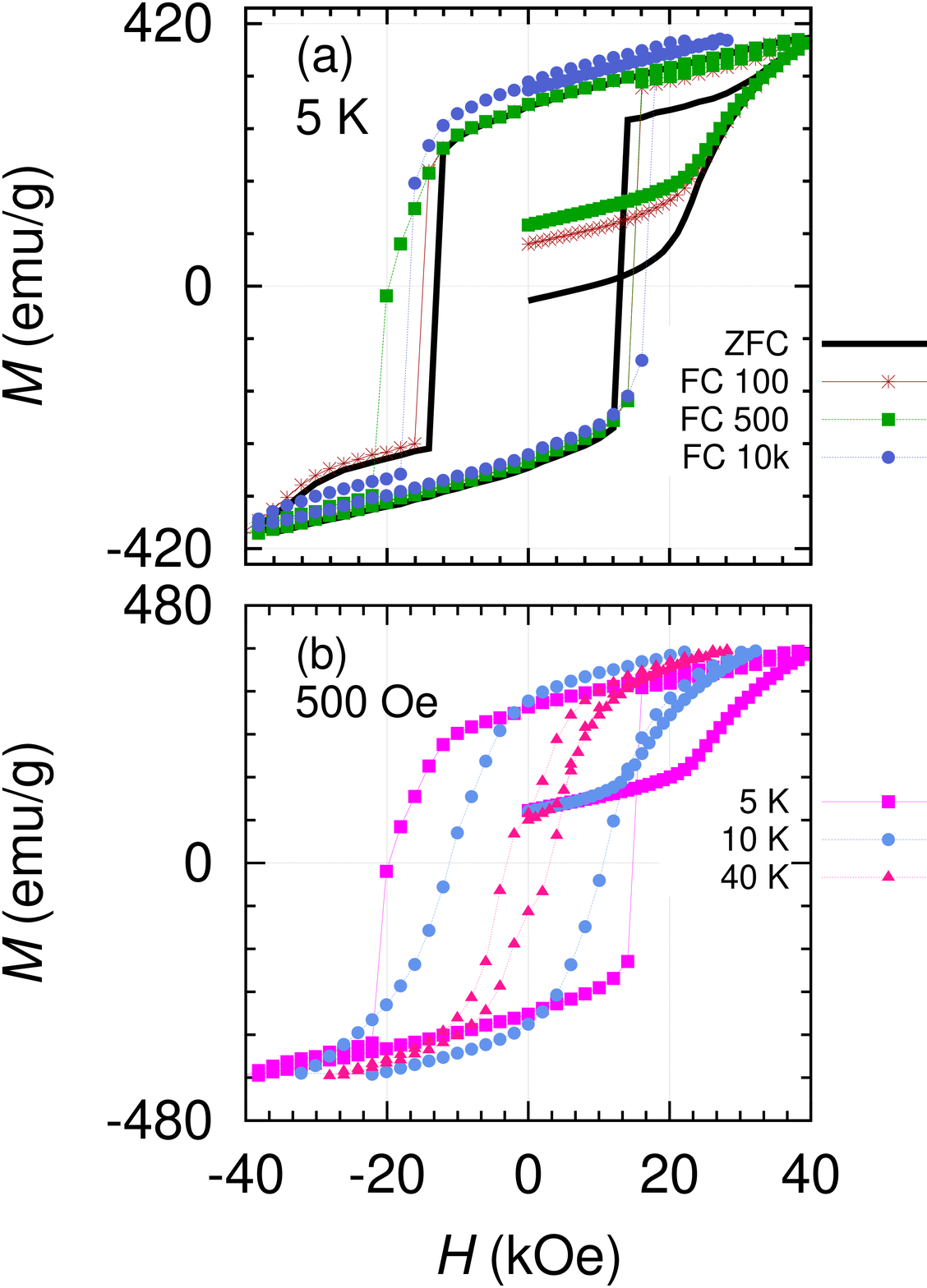}
\caption{\label{fig1_mh} (color online) (a) Magnetic hysteresis plots of Y$_2$CoMnO$_6$ at 5~K for zero field-cooled condition as well as different values of field-cooling. Shift of the FC-hysteresis loops compared to the ZFC is clear. (b) Hysteresis plots at different temperatures in the range 5 -- 40~K after field-cooling in 500~Oe. For the sake of clarity, not all the data measured are presented in the graphs.}
\end{figure}
\indent 
The samples used in the present investigation were prepared through solid state synthesis using stoichiometric ratios of Y$_2$O$_3$, MnO$_2$ and Co$_3$O$_4$ of high purity ($\mathrm{4N}$). The preliminary characterization of the samples was carried out using x rays, bulk magnetization measurements and detailed neutron diffraction studies are reported in [\onlinecite{nair_jap_2014}]. Our initial experiments on Y$_2$CoMnO$_6$ have shown the presence of prominent ferromagnetic hysteresis with metamagnetic "steps" below 8~K.\cite{nair_jap_2014} The neutron powder diffraction analysis quantified the {\it antisite} disorder in this material and confirmed the presence of magnetic inhomogeneity in the form of ferromagnetic and antiferromagnetic regions. We explained the presence of "steps" in magnetic hysteresis using a model for pinning of magnetization at the antiphase boundaries created by antisite disorder. These steps resembled the martensitic transformations found in intermetallics
and displayed first-order characteristics. Having understood that Y$_2$CoMnO$_6$ contains ferromagnetic and antiferromagnetic regions separated by antisite domain wall, it was obvious to test for exchange bias effect in this material. Hence, in the present work we carry out exchange bias experiments by measuring field-cooled and zero field-cooled magnetic hysteresis, {\it training effect} and ac susceptibility using a SQUID Magnetic Property Measurement System, Quantum Design Inc.
\\
\indent 
In Fig~\ref{fig1_mh} (a, b), the magnetic hysteresis measurements at 5~K performed in ZFC as well as FC cycles at different values of applied field and temperatures are presented.  The vertical and horizontal shifts in the FC magnetic hysteresis loops that are typical of EB systems are observed in Fig~\ref{fig1_mh} (a).  For each FC hysteresis measured at 5~K, the sample was prepared by heating it up to $T$ = 150~K $> T_c \approx$ 75~K to bring the sample to the paramagnetic state.  At this temperature, the magnetic field was removed in the "oscillate mode'' of the squid magnetometer.  The hysteresis curves in FC mode measured at different temperatures with applied field value of 500~Oe are presented in Fig~\ref{fig1_mh} (b).  The "steps" seen in the ZFC curve at 5~K\cite{nair_jap_2014} are smoothed in the field-cooled hysteresis curves.  It was observed in our previous study \cite{nair_jap_2014} that the "steps" depend on the sweep-rate of applied magnetic field.  The time- and sweep rate-dependent magnetic measurements suggested a domain structure for the material which was also supported by the evidence of mixed magnetic interactions and {\it antisite} disorder found through neutron scattering.  The magnetic hysteresis in Fig~\ref{fig1_mh} (b) shows weak step-like features, especially for the 40~K data. 
%
\begin{figure}[!t]
\centering{}
\includegraphics[scale=0.52]{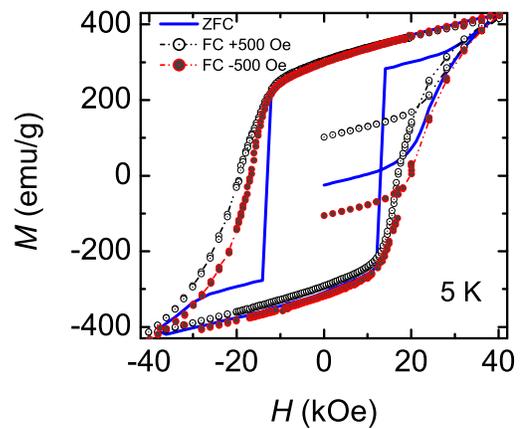}
\caption{\label{fig2_eb} (color online) Magnetic hysteresis in ZFC and FC conditions at 5~K clearly reveals the exchange bias shift. The FC curves are measured using +500~Oe and -500~Oe. The hysteresis loops shift in the opposite directions for positive and negative fields, respectively.}
\end{figure}
This already confirms the presence of multiple phases or an intrinsic inhomogeneity in the system. At this point we note a peculiarity in the exchange bias effect in Y$_2$CoMnO$_6$, that the EB shift in applied field of 10~kOe is slightly less than that at 500~Oe. The anisotropy field, $H_A$, was determined from initial magnetization curve at 5~K using the law of approach to saturation, $M$ = $M_s(1 - a/H - b/H^2)$ + $\chi$H and using the expressions, $b$ = 4$K^2_1$/15 $M^2_s$ and $H_A$ = 2$K_1$/$M_s$.\cite{patra_jpcm_21_078002_2009reply} $H_A \approx$ 90~Oe was estimated for Y$_2$CoMnO$_6$. This value is low in comparison with that of another double perovskite, Sr$_2$FeCoO$_6$ ($\approx$ 500~Oe)\cite{pradheesh_apl_101_142401_2012exchange} or from the case of Fe-doped LaMnO$_3$ ($\approx$ 20~kOe).\cite{patra_jpcm_21_078002_2009reply} We have measured hysteresis loops up to $H_{max}$ = 40~kOe $> H_A$. In order to confirm the EB effect, we performed field-cooled hysteresis measurements at 5~K using an applied field of +500~Oe as well as -500~Oe. This would cause the hysteresis loops to shift in opposite directions compared to the reference ZFC curve if exchange bias is present. Figure~\ref{fig2_eb}, hence, supports exchange bias in Y$_2$CoMnO$_6$.
\\
\indent 
The data from Fig~\ref{fig1_mh} (b) (which uses 500~Oe as the cooling field) was used to estimate the temperature dependence of exchange bias field, $H_{eb}$ = $(H_+ + H_-)/2$ (where $H_{\pm}$ are the positive and negative intercepts of the magnetization curve with the field axis) and the coercive field, $H_c$.  Figure~\ref{fig2_Heb}(a) presents the temperature evolution of $H_{c}$; the variation of $H_{eb}$ with temperature is shown in the Fig~\ref{fig2_Heb}(b).  One of the conclusive tests for exchange bias systems is the {\it{training effect}} which shows that $H_{eb}$ decreases with the number of consecutive loops of hysteresis measured.\cite{giri_jpcm_23_073201_2011exchange} 
Hence, we have measured the field cooled hysteresis curves at 5~K. A magnetic field of 60~kOe was used to field-cool the sample to 5~K before measuring hysteresis. Twenty loops of hysteresis were measured continuously. The dependence of $H_{eb}$ on the loop-number, $\lambda$, is shown in Fig~\ref{fig2_Heb} (c) based on the data from the measurement at 5~K. As expected for an exchange biased system, a gradual decrease of $H_{eb}$ upon increasing loop number is seen.
\\
\begin{figure}[!t]
\centering{}
\includegraphics[scale=0.22]{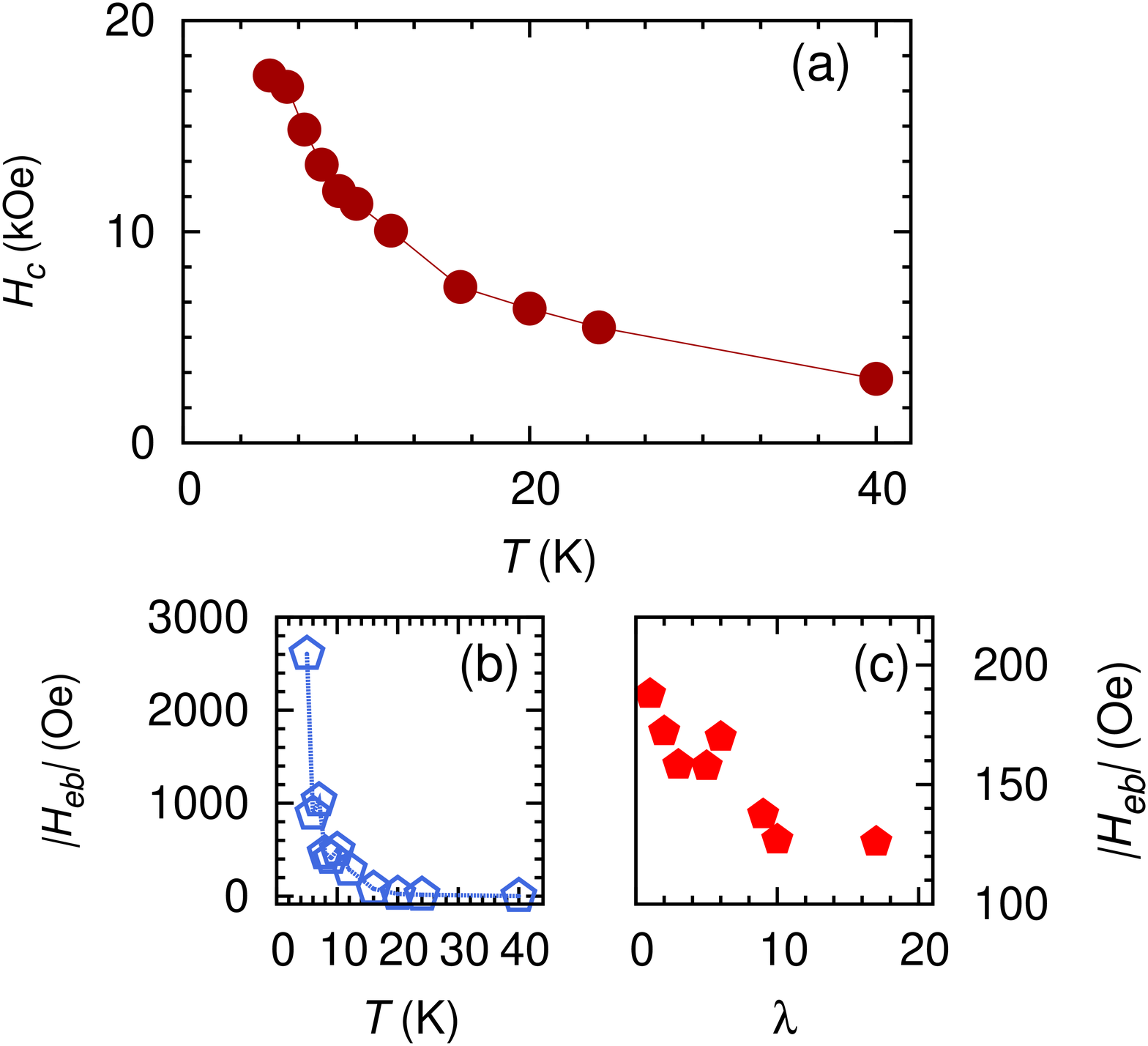}
\caption{\label{fig2_Heb} (color online) (a) Coercive field, $H_c$, versus temperature plot for Y$_2$CoMnO$_6$. Exchange bias field, $H_{eb}$, versus temperature estimated from hysteresis plots in Fig~\ref{fig1_mh} is shown in (b) while the dependence on $\lambda$, the number of loops measured in {\it{training effect}} experiment at 5~K is shown in (c).}
\end{figure}
\indent 
Exchange bias effect in complex bulk oxides result from interfaces between ferromagnetic, antiferromagnetic or spin-glass regions. Double perovskite compounds containing {\it antisite} defects have mixed magnetic exchange interactions that may lead to spin-glass-like phase which, in turn, can exhibit exchange bias through spin-glass interface between ferromagnetic or antiferromagnetic clusters.\cite{pradheesh_apl_101_142401_2012exchange} In order to probe the presence of such glassy phase in the case of Y$_2$CoMnO$_6$, ac susceptibility measurements were carried out at various frequencies ranging from 33~Hz to 6666~Hz and also by varying the amplitude of the ac field. The results are presented in Fig~\ref{fig3_acchi2} (a). The susceptibility peak at $T_c$ shows no frequency dependence in the measured range however, the imaginary parts of susceptibility show a "shoulder'' below $T_c$ which exhibits frequency-dependence (Fig~\ref{fig3_acchi2} (a)) and is also influenced by the increase in amplitude of the ac driving field. Ac susceptibility in cooling and warming cycles with an applied dc magnetic field was measured to check if the structurally multiple phases give rise to thermal hysteresis below $T_c$, but no effect was observed. As a final test for spin-glass-like phase in Y$_2$CoMnO$_6$, {\it{memory effect}} experiment was performed following the protocol reported in [\onlinecite{mathieu_prb_63_092401_2001}] and the results are shown in Fig~\ref{fig3_acchi2} (b). No signature of canonical spin-glass phase was observed. 
\\
\indent 
Neutron diffraction study on Y$_2$CoMnO$_6$ showed that the compound consists of disordered and ordered regions of monoclinic $P2_1/n$ and orthorhombic $Pnma$ structures at all temperatures in the range 4 -- 300~K.\cite{nair_jap_2014} The disordered regions in both the phases contain {\it antisite} defects which create antiferromagnetic superexchange pathways Mn$^{2+}$ -- O -- Mn$^{2+}$, Mn$^{4+}$ -- O -- Mn$^{4+}$ (similarly for Co) in addition to the Mn$^{2+}$ -- O -- Mn$^{4+}$ exchange which is ferromagnetic. This evidently leads to mixed magnetic exchange interactions and hence, cooling under magnetic field produces different magnetic domain structures than formed under zero field-cooling. The "steps" features in magnetization of Y$_2$CoMnO$_6$ can be accounted for by assuming that the domain walls separating ferromagnetic domains are trapped at the {\it antisite} regions and with high enough magnetic field, the pinning potential is overcome leading to an abrupt increase of magnetization. The scenario of pinning of domain walls with associated lattice strain has been recently modeled in the case of phase-separated manganites.\cite{niebieskikwiat_jpcm_24_436001_2012pinning} It is also interesting to note that through our neutron scattering experiments, we found that the volume fractions of the two structural phases did evolve with temperature.\cite{nair_jap_2014}
\\
\indent 
Ferromagnetic clusters in structurally disordered materials present strain fields that can pin the domain walls.  For example, it is understood that the antiphase boundaries in the double perovskites pin the magnetic domain walls.\cite{asaka_prb_75_184440_2007strong} The de-pinning of domain walls can then happen aided by applied magnetic field which creates surface excitations and lead to "bursts'' in magnetization.  The formalism of pinning of elastic objects described above is similar to the pinning of vortices in high-$T_c$ superconductors.\cite{larkin_31_784_1970effect} According to the model developed by Larkin, the surface tension of a typical domain wall is approximated as $\epsilon_t \sim 4\sqrt{\frac{JS^2K_1}{a}}$ where $J$ is the exchange constant between nearest neighbour spins of magnitude $S$, $K$ is the magnetocrystalline anisotropy energy per unit volume and $a$ is the distance between nearest neighbour spins.  The $J$ value for Y$_2$CoMnO$_6$ was approximated by the mean-field value $J_{MF}$ = 7.04~meV of closely related double perovskite material, La$_2$CoMnO$_6$. \cite{lv_jcc_33_1433_2012insulator} This leads to a value of $JS^2 \approx$ 1.69$\times$10$^{-21}$~J for Mn$^{4+}$ with $S$ = $\frac{3}{2}$.  A value of $K \approx$ 10$^5$ J/m$^3$ is adopted from the typical values observed in perovskites. With the nearest neighbour distance $a$ for Y$_2$CoMnO$_6$ as 0.395 nm as estimated from the previous structural analysis, an estimate of $\epsilon_t \sim$ 2.6~mJ/m$^2$ is obtained. In the formalism of Larkin, the amplitude of surface excitations follows the relation $\xi \sim \sqrt{\frac{U}{2\epsilon_t}}$. 
%
\begin{figure}[!t]
\includegraphics[scale=0.22]{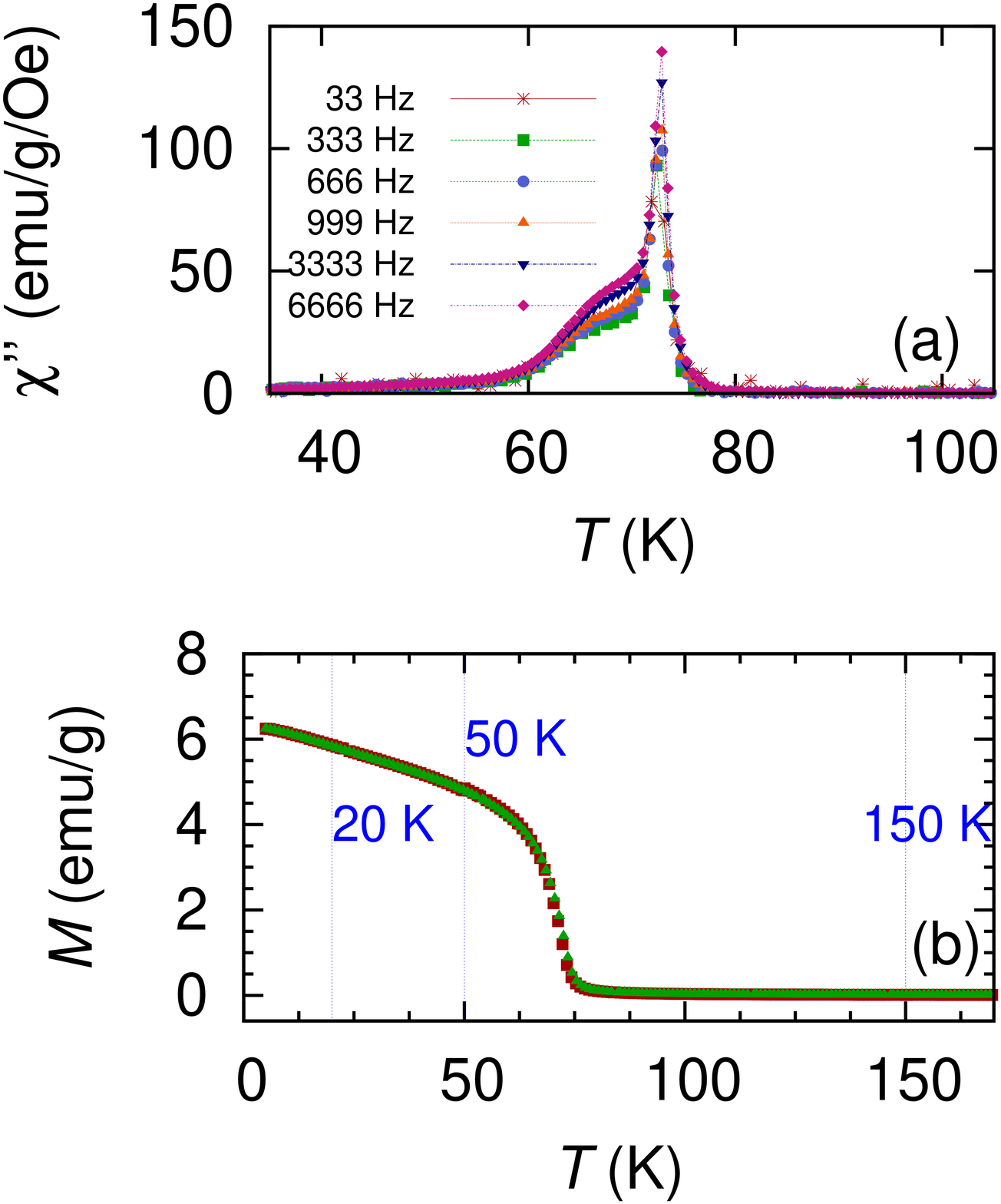}
\caption{(color online) (a) Imaginary part of ac susceptibility of Y$_2$CoMnO$_6$ at different frequencies. The real part of susceptibility shows no frequency dependence at $T_c \sim$ 75~K. A broad shoulder-like feature is visible in the imaginary part which appears to show dispersion. (b) Shows the {\it{memory effect}} test at 20, 50 and 150~K (marked by vertical dashed-lines) ruling out spin-glass phase. }
\label{fig3_acchi2}
\end{figure}
The ac susceptibility data could be described by thermally activated behaviour, $\chi(T)$ = $\chi_0$ + $\chi_1~ \mathrm{exp}(-U/k_BT)$ yielding a value, $U \sim$ 81.4~meV.  These values yield $\xi \approx$ 1.58~nm which is about four times the distance between nearest spins, $a$. Within the framework of pinning of domain walls, a scaling relation $\chi$ = ($\frac{3M_0}{H_C}$) $\times$ $\frac{\delta D_C}{D}$ is proposed.\cite{niebieskikwiat_jpcm_24_436001_2012pinning} 
Assuming the relation $\delta D_C \approx$ 2$\xi$ holds in the present case, we estimate the size of the ferromagnetic clusters as $D \sim$ 158~nm. Strong pinning of the magnetic domains at the antiphase boundaries have been experimentally directly observed in a double perovskite Ba$_2$FeMoO$_6$.\cite{asaka_prb_75_184440_2007strong} Microscopy methods combined with micromagnetic simulations can be an effective procedure to understand the details of the magnetic phases in the presence of {\it antisite} disorder. However in the present case, in order to correlate with the "steps" seen in hysteresis, experiments have to be performed in the presence of applied magnetic field. 
\\
Our experimental results combined with the observation of the {\em antisite} disorder in Y$_2$CoMnO$_6$ reported earlier\cite{nair_jap_2014}, suggest that the exchange bias in this material originates from {\em antisite} disorder. However, we wish to emphasis that we do not yet provide a "direct" link between the exchange bias and {\em antisite} disorder. A direct observation of the antiphase boundaries and domains and their evolution in applied magnetic field may lie within the reach of experimental observation using the state-of-the-art transmission electron microscopy, for example, the work by Asaka {\em et al.,}\cite{asaka_prb_75_184440_2007strong}. At this point we would like to point out that in order to understand this material completely several open questions remain to be answered, {\em viz.,} the role of sweep-rate on $H_{eb}$, the dependence of $H_{eb}$ on cooling field etc.
\\
\indent 
To conclude, the role of {\it antisite defects} in influencing magnetic properties of double perovskites  is clearly brought out through our observation of exchange bias in disordered Y$_2$CoMnO$_6$. Though the magnetic transition is observed at 75~K, the exchange bias is visible only below 8~K where the "steps'' in magnetic hysteresis occur. This clearly suggests a close connection between {\it antisite} disorder and the exchange bias effect. Through our ac susceptibility and {\it memory effect} experiments, we observe that the clusters do not lead to electronic phase separation or a spin-glass state. This narrows down the possible explanation of observed "steps'' in magnetic hysteresis to competition between elastic energy of the interfaces of domain walls and magnetic energy with applied field. The exchange bias is then suggested to arise at the interface of ferromagnetic clusters separated by antiferromagnetic antiphase domain boundaries.
%
%
%

%
%
%
\end{document}